\documentclass[journal,10pt,twocolumn,letter]{IEEEtran}

\usepackage{graphics}
\usepackage{colortbl}
\usepackage{multicol}
\usepackage{diagbox}
\usepackage{lipsum}
\usepackage{color}
\usepackage[cmex10]{amsmath}
\usepackage{amsthm}
\usepackage{amssymb}
\usepackage{mathrsfs}
\usepackage{mathtools}
\usepackage{amsbsy}
\usepackage[colorlinks=true,bookmarks=false,citecolor=blue,urlcolor=blue]{hyperref}
\usepackage{xcolor}
\usepackage{mathrsfs}
\usepackage{setspace}
\usepackage{float}
\usepackage{graphicx}
\usepackage{pstool}
\usepackage{lettrine}
\usepackage{newclude}
\usepackage[normalem]{ulem}
\usepackage{latexsym}
\usepackage{algpseudocode}
\usepackage{algorithm}
\usepackage{gensymb}

\usepackage{multirow}
\usepackage{amsmath}
\usepackage{amsmath,bm}
\usepackage{wasysym}
\usepackage{mathrsfs}
\usepackage{amsmath,cite,amsfonts,amssymb,color}
\usepackage{optidef}
\usepackage{dsfont}
\usepackage{bbm}

\ifCLASSOPTIONcompsoc
\usepackage[caption=false,font=normalsize,labelfont=sf,textfont=sf]{subfig}
\else
\usepackage[caption=false,font=footnotesize]{subfig}
\fi
\algdef{SE}[DOWHILE]{Do}{doWhile}{\algorithmicdo}[1]{\algorithmicwhile\ #1}

\allowdisplaybreaks

\graphicspath{{figures/}}
\allowdisplaybreaks

\newcommand{\RNum}[1]{\uppercase\expandafter{\romannumeral #1\relax}}

\usepackage{mathtools}

\begin{document}

\title{Performance Analysis of Uplink Optical Wireless Communications in Presence of a STAR-RIS}

\author{\IEEEauthorblockN{Alireza Salehiyan\thanks{Alireza Salehiyan and Mohammad Javad Emadi are with the Department of Electrical Engineering, Amirkabir University of Technology (Tehran Polytechnic), Tehran, Iran (E-mails: \{alirezasalehiyan,mjemadi\}@aut.ac.ir).} and Mohammad Javad Emadi
}}

% make the title area
\maketitle

%%====> Abstract <===%%
\begin{abstract}
Recently, reconfigurable intelligent surface (RIS) has gained research and development interests to modify wireless channel characteristics in order to improve performance of wireless communications, especially when quality of the line-of-sight channel is not that good. In this work, for the first time in the literature, we have used simultaneously transmitting and reflecting RIS (STAR-RIS) in non-orthogonal multiple-access visible light communication system to improve performance of the system. Achievable rates of the users are derived for two data recovery schemes, single-user detection (SUD) and successive interference cancellation (SIC). Then, sum-rate optimization problem is formulated for two operating modes of STAR-RIS, namely energy-splitting and mode-switching cases. Moreover, a sequential parametric convex approximation method is used to solve the sum-rate optimization problems. We have also compared energy-splitting and mode-switching cases and showed that these two modes have the same performance. Finally, numerical results for SUD and SIC schemes and two benchmarking schemes, time-sharing and max-min fairness, are presented and spectral- and energy-efficiency, number of STAR-RIS elements, position of users and access point are discussed.
\end{abstract}

\IEEEpeerreviewmaketitle

%%====> 1. Introduction <===%%
\section{Introduction}
	Beyond fifth-generation (5G) network is mainly supposed to serve massive number of connected devices, enhanced mobile broadband communications and low-latency applications, as well. Thus, to provide internet to everything, one of the main approaches is to improve the spectral efficiency by utilizing massive multiple antenna schemes, multi-user techniques such as non-orthogonal multiple access (NOMA) transmissions, successive interference cancelations (SIC), and using more bandwidth especially in millimeter-wave (mmWave) and terahertz (THz) communications. For the indoor communications, one of the promising approaches is to use optical wireless communications (OWC). Compared to the mmWave or sub-6 gigahertz communications, the light spectrum is un-licensed, it is secure since the light does not pass through the walls, there is no electromagnetic radiation concern, OWC can support higher data rate communications and is known as a green communications class. Therefore, OWC or visible light communication (VLC) systems with a large bandwidth of about 300 THz can be a promising candidate to be used in 6G, especially for indoor applications. It is worth noting that, similar to mmWave and THz communications, OWC systems are also sensitive to blockage and shadowing due to its short wavelength \cite{VincentRISVLCTutorial}. Moreover, it propagates in a line-of-sight (LOS) way and its performance dramatically degrades in non-line-of-sight (NLOS) scenarios.

	%In the recent years, with the growing number of internet users and the demand for higher data rates and bandwidth, the fifth-generation (5G) of cellular network was established and met the needs of users by using the millimeter-waves. However, in the future, as the millimeter-wave band becomes more congested, the available bandwidth will decrease again. Also, applications such as telemedicine and holographic communications require less latency than what is available on the 5G network. Hence, in recent years, researchers have been developing the sixth-generation (6G) network to meet the needs of users in the future.
	
%	Visible light communication (VLC) systems with a wide bandwidth of about 300 THz can be a good candidate for use in 6G. But, like millimeter-waves and terahertz waves, visible light is sensitive to blockage and shadowing due to its short wavelength \cite{VincentRISVLCTutorial}. Moreover, it propagates in a line-of-sight (LOS) way and performs poorly in non-line-of-sight (NLOS) scenarios.

To overcome the LoS blockage and constructively changing the channel environment to set-up a reliable communication, reconfigurable intelligent surface (RIS) is introduced recently to be used in both radio-frequency (RF) and optical wireless communications \cite{HassLiFi}. In the OWC, these surfaces use metasurface or mirror array to manipulate the incident wave received from a transmitter such that a desired destination can receive a high-power signal. To be more specific, the conventional RIS can absorb, amplify and reflect or transmit the impinging wave in an abnormal way \cite{HassLiFi}. Thus, they can be used to improve the spectral efficiency of the communication link. 

Because of the interesting properties of RISs, researchers have studied these surfaces from different perspectives in the literature. For instance in \cite{HassLCRIS} a structure based on liquid crystal is proposed to fabricate optical RISs. In \cite{AlouiniVLCRIS} authors have derived irradiance of photodetector in the presence of mirror array (MA)-based and metasurface-based RIS. Also, they have derived analytical expression for required phase gradient and reflector orientation in case of metasurface-based and MA-based RIS, respectively. In \cite{AlouiniChannel}, channel model of RIS-assisted VLC system along with its delay spread is derived. Moreover, authors in \cite{NdjiongueCapacity} have derived lower and upper bounds for capacity of VLC links with RIS, under average optical power and peak-intensity constraints. A diffused, NLOS VLC link is considered in \cite{PalaciosEvaluation} and impact of RIS on this link is evaluated. Additionally, a study on rate maximization for a single-user downlink VLC system with MA-based RIS and random user orientation and blockage is performed in \cite{NdjiongueIRS}. In \cite{SunMIMOVLC}, authors considered a downlink multiple-input multiple-output (MIMO) VLC system which uses RIS to minimize the mean square error (MSE) of demodulated signals at the receiver. Another study in \cite{SunSumrate} has considered a cell-free RIS-assisted VLC network with $L$ light-emitting diodes (LEDs) and $K$ users, where LEDs serve the users in time-division multiple access (TDMA) manner. Then, sum-rate of the system is maximized by using a greedy algorithm. Energy efficiency maximization of downlink RIS-aided VLC system is also studied in \cite{ChenReflecting} by jointly optimizing time allocation and power control coefficient and phase shift matrix. In \cite{QianSecure}, authors have derived the secrecy rate of downlink RIS-aided VLC system in presence of a legitimate user and an eavesdropper, and then maximized the rate to show how RIS can improve security of VLC links. For the first time in the literature, authors in \cite{hassNOMARIS2021} have used non-orthogonal multiple access (NOMA) techniques along with RIS in VLC links to improve link quality in presence of random user orientation and blockage, and minimize the maximum bit error rate (BER) of NOMA users. Also, In \cite{Hass2021DigitalR}, digital RIS is introduced, which controls the reflection of light in a digital manner.

Recently, a new variation of RIS called simultaneously transmitting and reflecting RIS (STAR-RIS) has been introduced in \cite{DobreSTAR,SchoberSTAR} to use in RF wireless systems. As its name implies, this surface can both reflect and transmit the impinging wave, thus, it can provide 360$\degree$ coverage. This type of RIS is gaining lots of attention in RF-based wireless communication. However, in optical wireless communication (OWC) systems, usage of STAR-RIS has not been investigated extensively. In \cite{Hass2022DoubleSidedBI}, for the first time in the literature, digital STAR-RIS is used in VLC systems, and fabrication method of this surface has been explained. Also, they have considered discrete phase shifts to have a fully digital STAR-RIS. 

In this paper, we consider a NOMA-based uplink STAR-RIS-assisted VLC link and seek to maximize the sum-rate of system by optimizing reflection and transmission coefficients of STAR-RIS to see the impact of this surface on the system performance. It is assumed that there are two rooms and one user is available in each room. Since there is only one AP in a given room, by using a STAR-RIS between the two rooms, the AP can receive the signal from the two users by utilizing the STAR-RIS. Firstly, we derive the users’ achievable rates with the assumption that the access point (AP) utilizes single-user detection or successive interference cancellation for data recovery. Then, we formulate the sum-rate optimization problem for two operating modes of STAR-RIS, that is energy-splitting and mode-switching schemes. Furthermore, to handle the non-convexity of the optimization problems, we use sequential parametric convex approximation (SPCA) algorithm to turn the problems into a convex one. Afterwards, we compare energy-splitting and mode-switching of STAR-RIS and show that they exhibit the same performance in terms of sum-rate. Finally, we simulate the two-user uplink OWC setup to verify our results through numerical methods. To the best of authors' knowledge, this is the first work which utilizes STAR-RIS in a NOMA-based OWC link.

\textit{Organization:} In Section \ref{Sec:Sec2}, system model of STAR-RIS-assisted VLC link is presented. The sum-rate maximization problem for two operation modes of STAR-RIS is formulated in Section \ref{Sec:Sec3} and solved in Section \ref{Sec:Sec4}. Moreover, Numerical results are presented in \ref{Sec:Sec5}. Finally, the paper is concluded in Section \ref{Sec:Sec6}.
%%====> 2. System Model <===%%
%%%%%%%%%%%%%%%%%%%%%%%%%%%%%%%%%%%%%%%%%%%%%%%%%%%%%%
\begin{figure}[t!]
    \centering
    \pstool[scale=0.6]{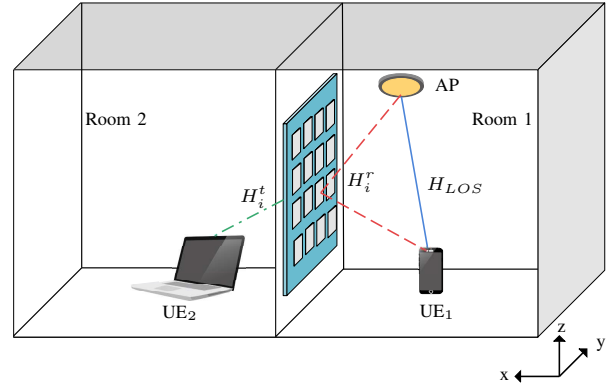}{
    \psfrag{1}{\hspace{-0.20cm}\scriptsize \text{Room 1}}
    \psfrag{2}{\hspace{-0.25cm}\scriptsize \text{Room 2}}
    \psfrag{A}{\hspace{-0.00cm}\scriptsize \text{$H_{LOS}$}}
    \psfrag{B}{\hspace{-0.00cm}\scriptsize \text{$H_i^r$}}
    \psfrag{C}{\hspace{-0.10cm}\scriptsize \text{$H_i^t$}}
    \psfrag{D}{\hspace{-0.00cm}\scriptsize \text{AP}}
    \psfrag{E}{\hspace{-0.10cm}\scriptsize \text{UE$_1$}}
    \psfrag{F}{\hspace{-0.10cm}\scriptsize \text{UE$_2$}}
    \psfrag{x}{\hspace{-0.00cm}\scriptsize \text{x}}
    \psfrag{y}{\hspace{-0.00cm}\scriptsize \text{y}}
    \psfrag{z}{\hspace{-0.00cm}\scriptsize \text{z}}
    }
    \caption{Two user uplink optical wireless communications by utilizing an STAR-RIS.}
    \label{Fig:Fig.1}
\end{figure}

%%%%%%%%%%%%%%%% Notations
\textit{Notations:} $ \boldsymbol{\beta} \in \mathbb{R}^{N\times 1} $ represents a real $N\times 1$ vector. Expected value of a random variable $x$ is denoted by $\mathbb{E}[x]$. A zero-mean Gaussian random variable $Z$ with variance $\sigma^2$ is denoted by $Z \sim \mathcal{N}(0,\sigma^2)$. Also,     $[x_1,x_2,...,x_N]$ represents a $1\times N$ vector with $x_1,x_2,...,x_N$ as its elements. Moreover, $\mathbbm{1}(.)$ and $\lVert \boldsymbol{x} \rVert$ stand for indicator function and norm of vector $\boldsymbol{x}$, respectively. Finally, $[0,1]$ denotes a set of numbers in range of 0 and 1, whereas, $\{0,1\}$ denotes a set of numbers which are either 0 or 1.

%%%%%%%%%%%%%%%%%%%%%%%%%%%%%%%%%%%%%%%%%%%%%%%%%%%%%%
\section{System Model} \label{Sec:Sec2}
\begin{figure}[t!]
    \centering
    \pstool[scale=0.8]{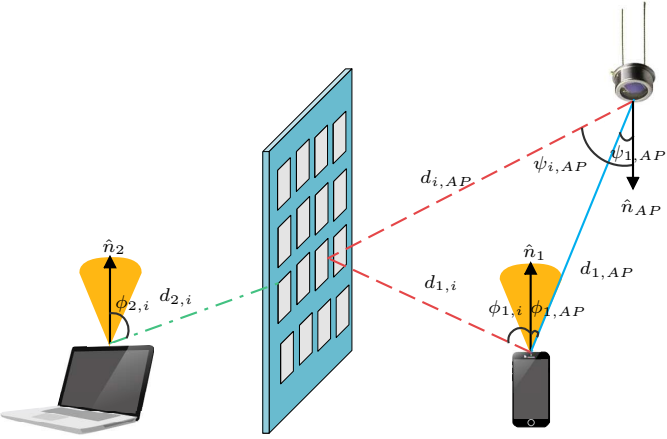}{
    \psfrag{A}{\hspace{0.00cm}\scriptsize \text{$d_{1,AP}$}}
    \psfrag{B}{\hspace{-0.10cm}\scriptsize \text{$d_{1,i}$}}
    \psfrag{C}{\hspace{-0.30cm}\scriptsize \text{$d_{i,AP}$}}
    \psfrag{E}{\hspace{-0.40cm}\scriptsize \text{$d_{2,i}$}}
    \psfrag{F}{\hspace{-0.10cm}\scriptsize \text{$\phi_{1,AP}$}}
    \psfrag{G}{\hspace{-0.25cm}\scriptsize \text{$\phi_{1,i}$}}
    \psfrag{I}{\hspace{-0.60cm}\scriptsize \text{$\psi_{i,AP}$}}
    \psfrag{J}{\hspace{-0.10cm}\scriptsize \text{$\psi_{1,AP}$}}
    \psfrag{K}{\hspace{-0.10cm}\scriptsize \text{$\phi_{2,i}$}}
    \psfrag{L}{\hspace{-0.05cm}\scriptsize \text{$\hat{n}_{1}$}}
    \psfrag{M}{\hspace{-0.05cm}\scriptsize \text{$\hat{n}_{AP}$}}
    \psfrag{N}{\hspace{-0.05cm}\scriptsize \text{$\hat{n}_{2}$}}
    }
    
    \caption{A channel geometry of signal transmission in presence of a STAR-RIS.}
    \label{Fig:Fig.2}
\end{figure}

As depicted in Fig.~\ref{Fig:Fig.1}, we have assumed an STAR-RIS with $N$ elements is located between two rooms 1 and  2, and there is an access point (AP) in room 1. In each room, there exists a user equipment (UE), wherein $k$-th UE for $k=1,2$, transmits its optical signal $x_k$ with optical power of $\mathbb{E}[x_k]=P_k$. Therefore, the received signal at the AP is given by
\begin{equation}\label{Eq:Eq1}
    y=\rho H_1x_1+\rho H_2x_2 + n,
\end{equation}
where $n \sim \mathcal{N}(0,\sigma_n^2)$ indicates the additive white Gaussian noise (AWGN), $\rho$ is the AP's photodetector's responsivity, and  effective channel coefficients $H_k$ are formulated as
\begin{subequations}
\begin{align}\label{Eq:Eq2}
        H_1& \triangleq H_{LOS}+\sum_{i=1}^N \beta_i^r H_i^r, \\
    H_2& \triangleq \sum_{i=1}^N \beta_i^t H_i^t,
\end{align}
\end{subequations}
where, $\beta_i^r,\beta_i^t$ $\in [0,1]$ denotes reflection and transmission coefficient of the $i$-th element of STAR-RIS, respectively, such that $\beta_i^r+\beta_i^t=1$ for $i=1,2,...,N$. Moreover, $H_{LOS}$ indicates the line-of-sight (LOS) channel coefficient between the UE in room 1 and the AP. Also, $H_i^r$ indicates the effective non-line-of-sight (NLOS) channel coefficient between UE$_{1}$, $i$-th element of RIS and AP, and $H_i^t$ is the the effective non-line-of-sight (NLOS) channel coefficient between UE$_{2}$, $i$-th element of RIS and AP, which are given as \cite{hassNOMARIS2021}
\begin{subequations}\label{Eq:Eq3}
\begin{multline}
        H_{LOS}=\frac{A_r(m+1)}{2\pi d_{1,AP}^2}\cos(\phi_{1,AP})^m \cos(\psi_{1,AP})G(\psi_{1,AP}) \\ \times \mathbbm{1}(0 \le \psi_{1,AP} \le \Psi_c),
\end{multline}

\begin{multline}
        H_i^r=\frac{A_r(m+1)}{2\pi (d_{1,i}+d_{i,AP})^2}\cos(\phi_{1,i})^m \cos(\psi_{i,AP})G(\psi_{i,AP}) \\ \times \mathbbm{1}(0 \le \psi_{i,AP} \le \Psi_c),
\end{multline}

\begin{multline}
        H_i^t=\frac{A_r(m+1)}{2\pi (d_{2,i}+d_{i,AP})^2}\cos(\phi_{2,i})^m \cos(\psi_{i,AP})G(\psi_{i,AP}) \\ \times \mathbbm{1}(0 \le \psi_{i,AP} \le \Psi_c),
\end{multline}
\end{subequations}\\
wherein $d_{1,AP}, d_{k,i}$ and $d_{i,AP}$ respectively denote the distance between (UE$_{1}$, AP), (UE$_{k}$, $i$-th element of RIS), and ($i$-th element of RIS, AP). Also, $A_r$ and $\Psi_c$ indicate the AP's photodetector's area and field of view (FOV), and $G(.)$ denotes the product of optical concentrator gain and filter gain. Also, $m$ is the Lambertian order of UEs' light source, that is \cite{ghassemlooy2019optical}

\begin{equation}\label{Eq:Eq4}
    m=-\frac{\ln(2)}{\ln(\cos(\Phi_{\frac{1}{2}}))},
\end{equation} 
where $\Phi_{\frac{1}{2}}$ is half-intensity angle of UEs' light source. Also, as illustrated in Fig.~\ref{Fig:Fig.2}, $\phi_{1,AP}, \psi_{1,AP}, \phi_{k,i}$, and $\psi_{i,AP}$ represent angle between ($\hat{n}_1, d_{1,AP}$), ($\hat{n}_{AP}, d_{1,AP}$), ($\hat{n}_k, d_{k,i}$) and ($\hat{n}_{AP},d_{i,AP}$) respectively, where $\hat{n}_k$ and $\hat{n}_{AP}$ are normal vectors of UE$_k$ and AP.
%%%%%%%%%%%%%%%%%%%%%%%%%%%%%%%%%%%%%%%%%%%%%%%%%%%%%%%
%%%%%%%%%%%%%%%%%%%%%%%%%%%%%%%%%%%%%%%%%%%%%%%%%%%%%%

By use of (\ref{Eq:Eq1}), the uplink achievable rate of $k$-th UE is derived as \cite{Wang2013} 
\begin{equation}\label{Eq:Eq5}
    R_k=\frac{1}{2}\log_2(1+\frac{e}{2\pi}\text{SINR}_k)~~~~ [\text{bpcu}].
\end{equation}
For the two well-know decoding schemes presented for multiple access channel \cite{el2011network}, i.e., single-user detection (SUD) and the successive interference cancellation (SIC) schemes, we have the two following signal to noise plus interference ratios (SINRs). For the SUD case, we have
\begin{equation}\label{Eq:Eq6}
    \text{SINR}_k=\frac{(\rho H_k P_k)^2}{\sigma_n^2+(\rho H_j P_j)^2},\enspace \text{for} \enspace k,j=1,2 , k \neq j.
\end{equation}
Also, SIC scheme wherein the AP firstly decodes user 2, and after canceling the 2nd user's interference, then decodes user 1, we have
\begin{subequations}\label{Eq:Eq7}
 \begin{equation}
    \text{SINR}_2=\frac{(\rho H_2 P_2)^2}{\sigma_n^2+(\rho H_1 P_1)^2}.
\end{equation}
 \begin{equation}
    \text{SINR}_1=\frac{(\rho H_1 P_1)^2}{\sigma_n^2},~~~~~~
 \end{equation}
 \end{subequations}

%%====> 3. Optimal Beamforming <===%%
\section{Sum-rate Maximization Problem} \label{Sec:Sec3}
In this section, the goal is to maximize the achievable uplink
sum-rate by optimizing the reflection and transmission coefficients of the STAR-RIS of size $N$. Two operating modes are considered for the STAR-RIS; \textit{energy-splitting} and \textit{mode-switching} schemes. In the following, the two optimization problems are defined.
\subsection{Energy-Splitting Mode}
For the energy-splitting mode, each element of the STAR-RIS can partially reflect and transmit the incident optical signal by splitting its energy \cite{SchoberSTAR2021}. Thus, the optimization problem is defined as
\begin{maxi!}
{\boldsymbol{\beta}^r, \boldsymbol{\beta}^t}{R_1+R_2 \label{Eq:P1Obj}}
{\label{Eq:P1}}{}
\addConstraint{0\le\beta^r_i\le 1, \forall i \label{Eq:C1P1}}
\addConstraint{0\le\beta^t_i\le 1, \forall i\label{Eq:C2P1}}
\addConstraint{\beta^r_i+\beta^t_i=1, \forall i, \label{Eq:C3P1}}
\end{maxi!}
where $\boldsymbol{\beta}^t \triangleq [\beta^t_1,\beta^t_1, ..., \beta^t_N]$, and $\boldsymbol{\beta}^r \triangleq [\beta^r_1,\beta^r_1, ..., \beta^r_N]$.
\subsection{Mode-Switching Mode}
In the mode-switching case, every element of the STAR-RIS either reflects or transmits the incident optical signal, i.e., $\beta^r_i$ and $\beta^t_i$ can be either 0 or 1. Therefore, for the mode-switching, the optimization problem becomes 
\begin{maxi!}
{\boldsymbol{\beta}^r, \boldsymbol{\beta}^t}{R_1+R_2 \label{Eq:P2Obj}}
{\label{Eq:P2}}{}
\addConstraint{\beta^r_i\in \{0,1\}, \forall i \label{Eq:C1P2}}
\addConstraint{\beta^t_i\in \{0,1\}, \forall i \label{Eq:C2P2}}
\addConstraint{\beta^r_i+\beta^t_i=1, \forall i. \label{Eq:C3P2}}
\end{maxi!}

According to (\ref{Eq:Eq5}), the optimization problem described in (\ref{Eq:P1}) is a non-convex one, and the one presented in (\ref{Eq:P2}) is a binary programming optimization problem  with a non-convex objective function. In next section, we have analytically solved the optimization problems by modifying them into convex forms.

%%====> 4. Analytical Solving of Beamforming Problems <===%%
\section{Optimizing Parameters of STAR-RIS}\label{Sec:Sec4}
To maximize the sum-rate of the two-user uplink data transmissions in presence of STAR-RIS, we firstly solve the optimization problems of energy-splitting case, and then the results are extended to the mode-switching scheme. 
%%%%%%%%%%%%%%%%%%%%%%%%%%%%%%%%%%%
\subsection{Energy-Splitting Scenario}
To solve the optimization problem given in (\ref{Eq:P1}), we firstly use the rates achieved by utilizing the SUD scheme presented in (\ref{Eq:Eq6}). We start by introducing auxiliary variables $u_k$ and rewrite the optimization problem (\ref{Eq:P1}) as follows,
\begin{maxi!}
{\boldsymbol{\beta}^r, \boldsymbol{\beta}^t , \boldsymbol{u}}{\sum_{k=1}^2 \frac{1}{2} \log_2(1+\frac{e u_k}{2\pi}) \label{Eq:P3Obj}}
{\label{Eq:P3}}{}
\addConstraint{ \frac{(\rho H_k P_k)^2}{\sigma_n^2+(\rho H_j P_j)^2} \ge u_k \enspace \text{for~~} k,j=1,2 \enspace k\neq j \label{Eq:C1P3}}
\addConstraint{\beta^r_i\in [0,1], \forall i \label{Eq:C2P3}}
\addConstraint{\beta^t_i\in [0,1], \forall i \label{Eq:C3P3}}
\addConstraint{\beta^r_i+\beta^t_i =1, \forall i. \label{Eq:C4P3}}
\end{maxi!}
Because of the non-convex constraint of (\ref{Eq:C1P3}), it is still hard to solve the modified optimization problem. Therefore to handle the non-convexity, we use the following auxiliary variables $v_j$ and rewrite (\ref{Eq:P3}) as
\begin{maxi!}
{\boldsymbol{\beta}^r, \boldsymbol{\beta}^t , \boldsymbol{u} ,  \boldsymbol{v}}{\sum_{k=1}^2 \frac{1}{2} \log_2(1+\frac{e u_k}{2\pi}) \label{Eq:P4Obj}}
{\label{Eq:P4}}{}
\addConstraint{(\rho H_k P_k)^2 \ge u_kv_j^2,  \enspace k,j=1,2 \enspace k \neq j                              \label{Eq:C1P4}}
\addConstraint{\sigma_n^2+(\rho H_j P_j)^2 \le v_j^2, \enspace j=1,2 \enspace j \neq k                            \label{Eq:C2P4}}
\addConstraint{\beta^r_i\in [0,1], \forall i\label{Eq:C3P4}}
\addConstraint{\beta^t_i\in [0,1], \forall i\label{Eq:C4P4}}
\addConstraint{\beta^r_i+\beta^t_i =1, \forall i. \label{Eq:C5P4}}
\end{maxi!}
Since $\rho , H_k$ and $P_k$ are positive, without loss of optimality,  (\ref{Eq:C1P4}) can be written as $\rho H_k P_k \ge \sqrt{u_k}v_j$. Also, (\ref{Eq:C2P4}) is equal to the following convex form
\begin{equation}\label{Eq:Eq8}
    \big\lVert\big[\rho H_j P_j,\sigma_n\big]\big\rVert \le v_j.
\end{equation}
Therefore, (\ref{Eq:P4}) is modified as
\begin{maxi!}
{\boldsymbol{\beta}^r, \boldsymbol{\beta}^t , \boldsymbol{u} ,  \boldsymbol{v}}{\sum_{k=1}^2 \frac{1}{2} \log_2(1+\frac{e u_k}{2\pi}) \label{Eq:P5Obj}}
{\label{Eq:P5}}{}
\addConstraint{\rho H_k P_k \ge \sqrt{u_k}v_j,  \enspace k,j=1,2 \enspace k \neq j                              \label{Eq:C1P5}}
\addConstraint{ \big\lVert\big[\rho H_j P_j,\sigma_n\big]\big\rVert \le v_j, \enspace j=1,2 \enspace j \neq k                            \label{Eq:C2P5}}
\addConstraint{\beta^r_i\in [0,1], \forall i\label{Eq:C3P5}}
\addConstraint{\beta^t_i\in [0,1], \forall i\label{Eq:C4P5}}
\addConstraint{\beta^r_i+\beta^t_i =1, \forall i.\label{Eq:C5P5}}
\end{maxi!}
Still, due to (\ref{Eq:C1P5}) the problem is non-convex. To deal with non-convexity, SPCA method is used \cite{beck2009}, \cite{ShenRateMaximization}, in which the term $\sqrt{u_k}v_j$ is replaced with its convex upper bound, $\frac{u_k}{2\theta_k}+\frac{v_j^2\theta_k}{2}$, with equality when $\theta_k=\frac{\sqrt{u_k}}{v_j}$. 
Therefore, (\ref{Eq:C1P5}) can be rewritten as $\rho H_k P_k \ge \frac{u_k}{2\theta_k}+\frac{v_j^2\theta_k}{2}$ which is convex, and values of $\theta_k, u_k, v_j ,\beta^r_i$ and $\beta^t_i$ are updated iteratively until convergence. In the other words, at the $m$-th iteration of SPCA, the following problem is solved
\begin{maxi!}
{\boldsymbol{\beta}^r, \boldsymbol{\beta}^t , \boldsymbol{u} ,  \boldsymbol{v}}{\sum_{k=1}^2 \frac{1}{2} \log_2(1+\frac{e u_k}{2\pi}) \label{Eq:P6Obj}}
{\label{Eq:P6}}{}
\addConstraint{\rho H_k P_k \ge \frac{u_k}{2\theta_k^{(m)}}+\frac{v_j^2\theta_k^{(m)}}{2},  \enspace k,j=1,2 \enspace k \neq j                              \label{Eq:C1P6}}
\addConstraint{ \big\lVert\big[\rho H_j P_j,\sigma_n\big]\big\rVert \le v_j, \enspace j=1,2 \enspace j \neq k                            \label{Eq:C2P6}}
\addConstraint{\beta^r_i\in [0,1], \forall i\label{Eq:C3P6}}
\addConstraint{\beta^t_i\in [0,1], \forall i\label{Eq:C4P6}}
\addConstraint{\beta^r_i+\beta^t_i =1, \forall i. \label{Eq:C5P6}}
\end{maxi!}
where $\theta_k^{(m)}=\frac{\sqrt{u_k^{(m-1)}}}{v_j^{(m-1)}}$, and $u_k^{(m-1)},v_j^{(m-1)}$ are   solutions of the last iteration. The SPCA algorithm for solving (\ref{Eq:P1}) is summarized in \textbf{Algorithm}~\ref{Alg:Alg.1}.

Following the same procedure, the optimization problem (\ref{Eq:P1}) can be modified as follows to consider the achievable sum-rate of the SIC scheme. 
\begin{maxi!}
{\boldsymbol{\beta}^r, \boldsymbol{\beta}^t , \boldsymbol{u} ,\boldsymbol{v}}{\sum_{k=1}^2 \frac{1}{2} \log_2(1+\frac{e u_k}{2\pi}) \label{Eq:P7Obj}}
{\label{Eq:P7}}{}
\addConstraint{\frac{\rho H_1 P_1}{\sigma_n} \ge \frac{u_1}{2\theta_1^{(m)}}+\frac{\theta_1^{(m)}}{2} \label{Eq:C1P7}}
\addConstraint{\rho H_2 P_2 \ge \frac{u_2}{2\theta_2^{(m)}}+\frac{v^2\theta_2^{(m)}}{2} \label{Eq:C2P7}}
\addConstraint{ \big\lVert\big[\rho H_1 P_1,\sigma_n\big]\big\rVert \le v \label{Eq:C3P7}}
\addConstraint{\beta^r_i\in [0,1], \forall i\label{Eq:C4P7}}
\addConstraint{\beta^t_i\in [0,1], \forall i\label{Eq:C5P7}}
\addConstraint{\beta^r_i+\beta^t_i =1, \forall i. \label{Eq:C6P7}}
\end{maxi!}
%%%%%%%%%%%%%%%%%%%%%%%%%%
\subsection{Mode-Switching Scenario}
The optimization problem described in (\ref{Eq:P2}), is a non-convex binary-programming problem which is hard to solve analytically. However, for now, lets consider a single element RIS with reflection and transmission coefficients of $\beta^r$ and $\beta^t=1-\beta^r$ respectively, and take a deeper look at the objective function, which can be rewritten by inserting (\ref{Eq:Eq5}) and (\ref{Eq:Eq6}) into (\ref{Eq:P1Obj}) as follows
\begin{equation}\label{Eq:Mode_Switching1}
\begin{split}
f(\beta^r)&= \frac{1}{2}\log_2\Bigg(1+\frac{e}{2\pi}\frac{\rho^2(H_{LOS}+H^r\beta^r)^2P_1^2}{\rho^2{H^t}^2(1-\beta^r)^2P_2^2+\sigma_n^2}\Bigg) \\ 
&+ \frac{1}{2}\log_2\Bigg(1+\frac{e}{2\pi}\frac{\rho^2{H^t}^2(1-\beta^r)^2P_2^2}{\rho^2(H_{LOS}+H^r\beta^r)^2P_1^2+\sigma_n^2}\Bigg).
\end{split}
\end{equation}
For this objective function, we have
\begin{subequations}\label{Eq:Mode_Switching2}
 \begin{equation}
 \begin{split}
     f(0)&=\frac{1}{2}\log_2\Bigg(1+\frac{e}{2\pi}\frac{\rho^2 H_{LOS}^2P_1^2}{\rho^2{H^t}^2P_2^2+\sigma_n^2}\Bigg) \\ 
      &+ \frac{1}{2}\log_2\Bigg(1+\frac{e}{2\pi}\frac{\rho^2{H^t}^2P_2^2}{\rho^2H_{LOS}^2P_1^2+\sigma_n^2}\Bigg),
 \end{split}
 \end{equation}
 
 \begin{equation}
 \quad f(1)=\frac{1}{2}\log_2\Bigg(1+\frac{e}{2\pi}\frac{\rho^2(H_{LOS}+H^r)^2P_1^2}{\sigma_n^2}\Bigg).
 \end{equation}
\end{subequations}
Since $0\le\beta^r\le1 $, it can be deduced from (\ref{Eq:Mode_Switching1}) and (\ref{Eq:Mode_Switching2}) that $\max f(\beta^r)=\max\big(f(0),f(1)\big)$, i.e, maximum of objective function occurs either at 0 or 1. This relation also holds for $N$-dimensional objective function. Consequently, performance of mode-switching is the same as energy-splitting and there is no need to solve (\ref{Eq:P2}) since it yields the same result as (\ref{Eq:P1}). Anyway, for analytical solving of (\ref{Eq:P2}), one can use penalty method described in \cite{STARSumrate}.

\begin{algorithm}[t!]
\caption{SPCA Algorithm for Solving (\ref{Eq:P1})}\label{Alg:Alg.1}
\begin{algorithmic}[1]
\State \textit{Initialization:} set $\theta_k^{(0)}$, iteration index $m=0$ and convergence accuracy $\epsilon$.
\Do
\State Solve the convex problem in (\ref{Eq:P6}).
\State Set $\theta_k^{(m+1)}=\frac{\sqrt{u_k^{(m)}}}{v_j^{(m)}}$.
\State $m=m+1$.
\doWhile{Difference of all variables from their last iteration's values are less than $\epsilon$}
\end{algorithmic}
\end{algorithm}
%%====> 5. Numerical Results and Discussions <===%%
\section{Numerical Results and Discussions} \label{Sec:Sec5}
In this section, performance of STAR-RIS-assisted VLC system is analysed through numerical results, wherein energy efficiency and sum-rate are used as performance metric. Also the two operation modes of STAR-RIS are compared and convergence of SPCA algorithm is also discussed.  
%%%%%%%%%%%%%%%%%%%%%%
\subsection{Setup Parameters}
To simulate the channel coefficients, two rooms 1 and 2, each with the same dimension $5 \times 5 \times 3 \, m^3$ are assumed. An access point is located at the ceiling of room 1 with position of $[x,y,z]=[4.5,2.5,3] \, m$. UE$_1$ is in room 1 and located at $[x,y,z]=[3.5,2.5,3] \, m$ and UE$_2$ is in room 2 and its location is $[x,y,z]=[6,2.5,3] \, m$. Moreover, an STAR-RIS is placed between the two rooms, with center position at $[x,y,z]=[5,2.5,1.5] \, m$. Other parameters are listed in Table \ref{Tab:Tab1}.
%%=============================================================%%
\begin{table}[t!]
\centering
\caption{Simulation parameters.}\label{Tab:Tab1}
\begin{tabular}[t]{lc}
\hline
Parameter&Value\\
\hline
\hline
Rooms' Dimension&$5\times 5 \times 3\, m^3$\\
Access Point's Location&$[4.5,2.5,3]\,m$\\
UE$_1$'s Location&$[3.5,2.5,1]\,m$\\
UE$_2$'s Location&$[6,2.5,1]\,m$\\
STAR-RIS Center's Location&$[5,2.5,1.5]\,m$\\
STAR-RIS's Size&$10\times8$ elements\\
LED's Half-Angle&$60\degree$\\
Photodetector's Half-Angle&$85\degree$\\
Photodetector's Area&$1.5\,cm^2$\\
Photodetector's Responsivity&$0.7$\\
Gain of Optical Concentrator and Filter&$10$\\
UE$_1$'s Power&$100\,mW$\\
UE$_2$'s Power&$100\,mW$\\
Noise Variance&$10^{-10}\, \frac{W}{Hz}$\\
\hline
\end{tabular}
\end{table}
%%=============================================================%%
\subsection{Numerical Results}
In what follows, we consider the energy splitting and mode-switching cases, and convergence of the iterative algorithm.

\subsubsection{Energy Splitting}
\begin{figure}[t]
    \centering
    \pstool[scale=0.55]{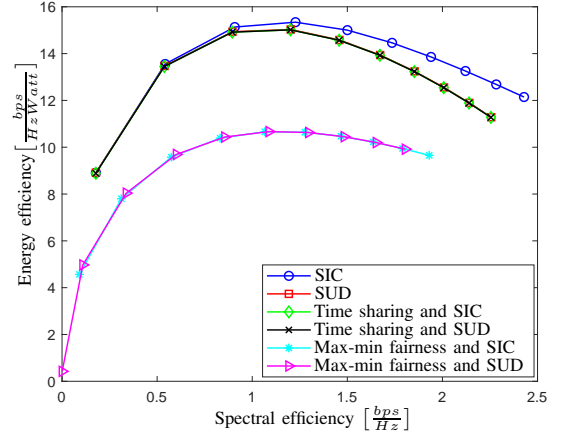}{
    \psfrag{SE}{\hspace{-0.7cm}\scriptsize \text{Spectral efficiency $\big[\frac{bps}{Hz}\big]$}}
    \psfrag{EE}{\hspace{-0.00cm}\scriptsize \text{Energy efficiency$\big[\frac{bps}{Hz Watt}\big]$}}
    \psfrag{SIC}{\hspace{-0.00cm}\scriptsize \text{SIC}}
    \psfrag{SUD}{\hspace{-0.00cm}\scriptsize \text{SUD}}
    \psfrag{ABCDEFGHIJKLMNOP}{\hspace{-0.00cm}\scriptsize \text{Time sharing and SIC}}
    \psfrag{ABCDEFGHIJKLMNOPQ}{\hspace{-0.00cm}\scriptsize \text{Time sharing and SUD}}
    \psfrag{ABCDEFGHIJKLMNOPQRSTUV}{\hspace{-0.00cm}\scriptsize \text{Max-min fairness and SIC}}
    \psfrag{ABCDEFGHIJKLMNOPQRSTUVWX}{\hspace{-0.00cm}\scriptsize \text{Max-min fairness and SUD}}
    }
    
    \caption{Energy-- versus spectral--efficiency for UE's power from 0 to $100 mW$.}
    \label{Fig:Fig.3}
\end{figure}

The energy efficiency versus spectral efficiency of the energy splitting case for different methods are discussed in Fig.~\ref{Fig:Fig.3}. The energy efficiency metric is defined as
\begin{equation}\label{Eq:Eq10}
\eta_{EE}\triangleq \frac{R_1+R_2}{P_1+P_2}.
\end{equation}
Besides the SUD and SIC schemes, for bench marking, we have also considered the two well-known cases; the max-min and the time-sharing approaches\footnote{\textcolor{black}{For the max-min fairness and the time sharing scheme, objective function of (\ref{Eq:P1}) respectively changes to $\underset{\boldsymbol{\beta^r},\boldsymbol{\beta^t}}{\text{{maximize}}}\enspace \text{{minimum}}\,(R_1,R_2)$, and $\underset{\boldsymbol{\beta^r},\boldsymbol{\beta^t},\alpha}{\text{{maximize}}}\enspace \alpha R_1+(1-\alpha) R_2$. With the same constraints as (\ref{Eq:P1}) in addition to $0\le \alpha \le1$ for time sharing.}}. \\
As depicted in Fig.~\ref{Fig:Fig.3}, by increasing UEs' power from $0$ to $100\,mW$, both spectral efficiency and energy efficiency start to increase. However, after one point, energy efficiency decreases gradually while spectral efficiency keeps increasing. This decrease in energy efficiency is more significant in SUD and time sharing. Also SUD and time sharing performances are the same. Moreover, max-min shows poorest efficiency in comparison with other methods. Therefore, in the rest of discussion, only SIC and SUD are discussed.
%%=============================================================%%
%===>SE-EE<===%
%%=============================================================%%

\begin{figure}[t]
    \centering
    \pstool[scale=0.55]{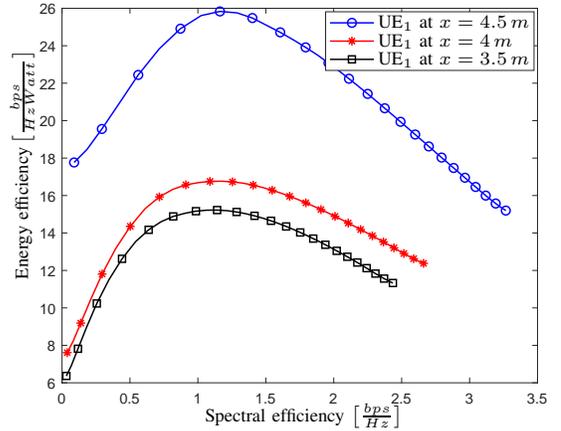}{
    \psfrag{SE}{\hspace{-0.7cm}\scriptsize \text{Spectral efficiency $\big[\frac{bps}{Hz}\big]$}}
    \psfrag{EE}{\hspace{-0.00cm}\scriptsize \text{Energy efficiency\,$\big[\frac{bps}{Hz Watt}\big]$}}
    \psfrag{UE1A1234567891011121}{\hspace{-0.0cm}\scriptsize \text{UE$_1$ at $x=4.5\,m$ }}
    \psfrag{UE1B1234567891011121}{\hspace{-0.0cm}\scriptsize \text{UE$_1$ at $x=4\,m$ }}
    \psfrag{UE1C1234567891011121}{\hspace{-0.0cm}\scriptsize \text{UE$_1$ at $x=3.5\,m$ }}
    }
    \caption{Energy-- versus spectral-efficiency of SIC for different position of UE$_1$.}
    \label{Fig:Fig.4}
\end{figure}

Energy efficiency of SIC versus spectral efficiency is depicted in Fig.~\ref{Fig:Fig.4} for different distance of UE$_1$ from RIS. The more UE$_1$ gets closer to AP and RIS, the more energy efficiency is increased. When UE$_1$ is placed at $x=4.5\,m$, it has the least distance with AP and RIS, therefore energy-- and spectral--efficiency increase significantly.
%%=============================================================%%
%===>SE-EE SIC<===%
%%=============================================================%%

\begin{figure}[t]
    \centering
    \pstool[scale=0.55]{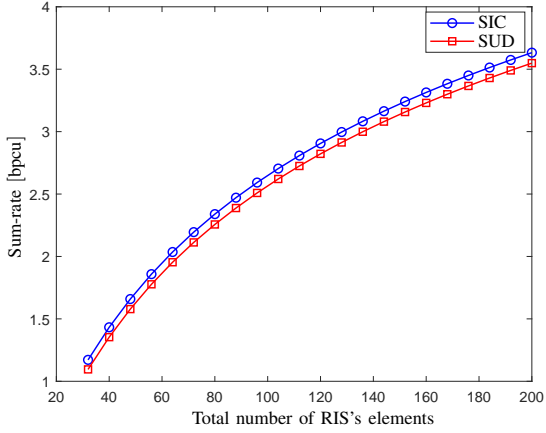}{
    \psfrag{SIC}{\hspace{-0.00cm}\scriptsize \text{SIC}}
    \psfrag{SUD12}{\hspace{-0.00cm}\scriptsize \text{SUD}}
    \psfrag{ABCDEFGHI}{\hspace{-0.2cm}\scriptsize \text{Total number of RIS's elements}}
    \psfrag{SumRate}{\hspace{-0.00cm}\scriptsize \text{Sum-rate $[\text{bpcu}]$}}
    }
    \caption{Sum-rate-- versus total number of RIS's elements.}
    \label{Fig:Fig.5}
\end{figure}

In Fig.~\ref{Fig:Fig.5}, sum-rate versus total number of RIS's elements is plotted for both SIC and SUD. As the number of RIS's elements increases, more elements are available for reflecting and transmitting the optical signal, and thus, the sum-rate increases too. However, there is a gap between sum-rate of SIC and SUD, because in SUD, signal of one user at the AP is always treated as noise for another user's signal and causes a decrease in sum-rate.
%%=============================================================%%
%===>Rate-Size<===%
%%=============================================================%%

\begin{figure}[t!]
    \centering

    \pstool[scale=0.55]{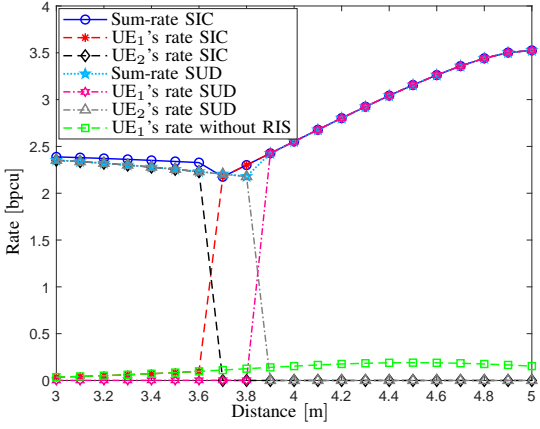}{
    \psfrag{SumRateSUD}{\hspace{-0.00cm}\scriptsize \text{Sum-rate SUD}}
    \psfrag{Rate1SUD}{\hspace{-0.00cm}\scriptsize \text{UE$_1$'s rate SUD}}
    \psfrag{Rate2SUD}{\hspace{-0.00cm}\scriptsize \text{UE$_2$'s rate SUD}}
    \psfrag{SumRateSIC}{\hspace{-0.00cm}\scriptsize \text{Sum-rate SIC}}
    \psfrag{Rate1SIC}{\hspace{-0.00cm}\scriptsize \text{UE$_1$'s rate SIC}}
    \psfrag{Rate2SIC}{\hspace{-0.00cm}\scriptsize \text{UE$_2$'s rate SIC}}
    \psfrag{Rate1WithoutRIS12345678910}{\hspace{-0.00cm}\scriptsize \text{UE$_1$'s rate without RIS}}
    \psfrag{Rate}{\hspace{-0.00cm}\scriptsize \text{Rate $[\text{bpcu}]$}}
    \psfrag{Distance}{\hspace{-0.35cm}\scriptsize \text{Distance $[\text{m}]$}}
    }
    \caption{User rates-- versus UE$_1$'s $x$ position in room 1. UE$_2$ is fixed at $x=6\,m$.}
    \label{Fig:Fig.6}
\end{figure}

\begin{figure}[ht]
    \centering
    \pstool[scale=0.55]{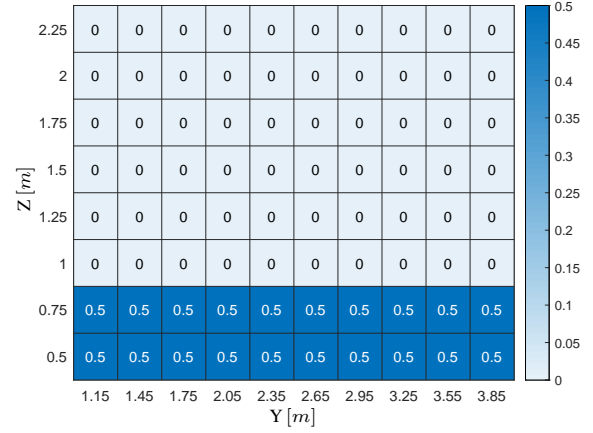}{
    \psfrag{Z123}{\hspace{-0.0cm}\scriptsize \text{Z\,$[m]$}}
    \psfrag{Y}{\hspace{-0.25cm}\scriptsize \text{Y\,$[m]$}}
    }
    \caption{Reshaped matrix of reflection coefficients ($\boldsymbol{\beta^r}$) vector of RIS for UE$_1$ at $x=3.5\,m$ and UE$_2$ at $x=6\,m$.}
    \label{Fig:Fig.7}
\end{figure}

The achievable rates of users versus position of UE$_1$ for SIC and SUD schemes are depicted in Fig.~\ref{Fig:Fig.6}. As UE$_1$ approaches AP from $x=3\,m$ to $x=3.6\,m$, UE$_2$'s rate decreases slightly, while rate of UE$_1$ increases. At this stage, the increase in UE$_1$'s rate is because of LOS path that exists between AP and UE$_1$. In other words, RIS is mainly transmitting UE$_2$ signal and reflects a very small portion of UE$_1$'s signal as depicted in Fig.~\ref{Fig:Fig.7}. After $x=3.6\,m$, since line-of-sight path between UE$_1$ and AP has become better, RIS mainly reflects UE$_1$'s signal and ignores UE$_2$. This trend also happens for SUD, but with a difference that rate due to LOS path is substantially reduced because UE$_2$'s signal is treated as noise for UE$_1$.
%%=============================================================%%
%===>Rate-Distance: UE2 fixed<===%
%%=============================================================%%
%===>Beta_t r UE1 at x=3.5<===%
%%=============================================================%%

\begin{figure}[t]
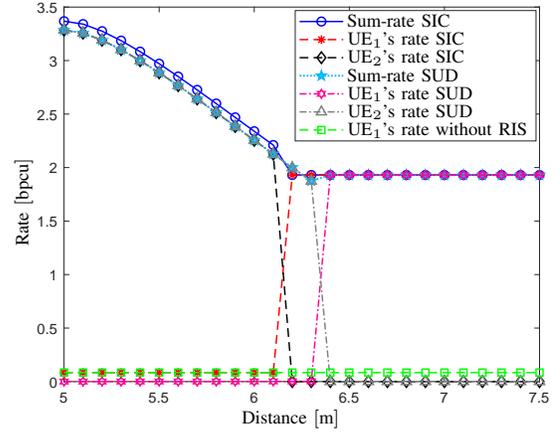

    \centering
    \pstool[scale=0.55]{Result/Rate_Distance_UE1_Fixed_Both_Techniques}
    {
    \psfrag{SumRateSUD}{\hspace{-0.00cm}\scriptsize \text{Sum-rate SUD}}
    \psfrag{Rate1SUD}{\hspace{-0.00cm}\scriptsize \text{UE$_1$'s rate SUD}}
    \psfrag{Rate2SUD}{\hspace{-0.00cm}\scriptsize \text{UE$_2$'s rate SUD}}
    \psfrag{SumRateSIC}{\hspace{-0.00cm}\scriptsize \text{Sum-rate SIC}}
    \psfrag{Rate1SIC}{\hspace{-0.00cm}\scriptsize \text{UE$_1$'s rate SIC}}
    \psfrag{Rate2SIC}{\hspace{-0.00cm}\scriptsize \text{UE$_2$'s rate SIC}}
    \psfrag{Rate1Without RIS12345678910}{\hspace{-0.00cm}\scriptsize \text{UE$_1$'s rate without RIS}}
    \psfrag{Rate}{\hspace{-0.00cm}\scriptsize \text{Rate $[\text{bpcu}]$}}
    \psfrag{Distance}{\hspace{-0.35cm}\scriptsize \text{Distance $[\text{m}]$}}
    }
    \caption{User rates-- versus UE$_2$'s $x$ position in room 2. UE$_1$ is fixed at $x=3.5\,m$.}
    \label{Fig:Fig.9}
\end{figure}

Now, in Fig.~\ref{Fig:Fig.9}, UE$_1$ is fixed at $x=3.5\,m$ and user rates versus UE$_2$'s position, i.e. $x$, is plotted. When UE$_2$ is near RIS, most elements transmit UE$_2$ signal. As it get far from the RIS, UE$_2$'s rate starts to fall off. This trend continues up to $x=6.1\,m$, where, the signal that reaches to AP from UE$_2$ is so weak that RIS prefers to mostly reflect the signal of UE$_1$ to maintain the sum-rate.
%%=============================================================%%
%===>Rate-Distance: UE1 fixed<===%

\begin{figure}[ht]
    \centering
    \pstool[scale=0.55]{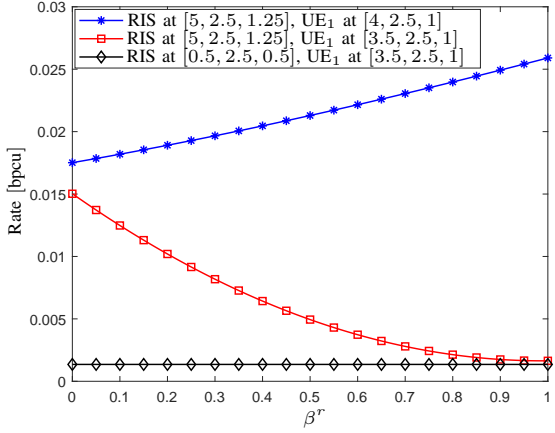}{
    \psfrag{Br}{\hspace{-0.0cm}\scriptsize \text{$\beta^r$}}
    \psfrag{Rate}{\hspace{-0.00cm}\scriptsize \text{Rate $[\text{bpcu}]$}}
    \psfrag{1ABCDEFGHIJKLMNOPQRSTUVWXYZABCDEFGHIJKLM}{\hspace{-0.0cm}\scriptsize \text{RIS at $[5,2.5,1.25]$, UE$_1$ at $[4,2.5,1]$ }}
    \psfrag{0ABCDEFGHIJKLMNOPQRSTUVWXYZ}{\hspace{-0.0cm}\scriptsize \text{RIS at $[5,2.5,1.25]$, UE$_1$ at $[3.5,2.5,1]$ }}
    \psfrag{5ABCDEFGHIJKLMNOPQRSTUVWXYZ}{\hspace{-0.0cm}\scriptsize \text{RIS at $[0.5,2.5,0.5]$, UE$_1$ at $[3.5,2.5,1]$ }}
    }
    \caption{Sum-rate-- versus $\beta^r$.}
   \label{Fig:Fig.13}
\end{figure}

It should be noted that in Fig.~\ref{Fig:Fig.7}, $\boldsymbol{\beta^r}=0.5$ has no effect on sum-rate. To clarify, sum-rate is plotted versus $\beta^r$ of a \textit{single element} RIS for different positions of element and UE$_1$ in Fig.~\ref{Fig:Fig.13}. When RIS element is located at $[5,2.5,1.25]$, it is accessible by both users and can reflect or transmit depending on distance of UEs from RIS. Moreover, in this case optimum point always occur at $\beta^r=0$ or $\beta^r=1$. However, when RIS element is located at $[5,2.5,0.5]$, neither UE$_1$, nor UE$_2$ cannot access this element and their signals do not even reach RIS. Therefore, this element is useless and does not change the rate, thus it doesn't matter what value it takes, all values in $[0,1]$ are optimum for this element. So, it is better to disable such elements that are not accessible by users and turn them off.

%%=============================================================%%
%========>Optimum Beta<========%
%%=============================================================%%

\begin{figure}[t]
 \centering
    \pstool[scale=0.65]{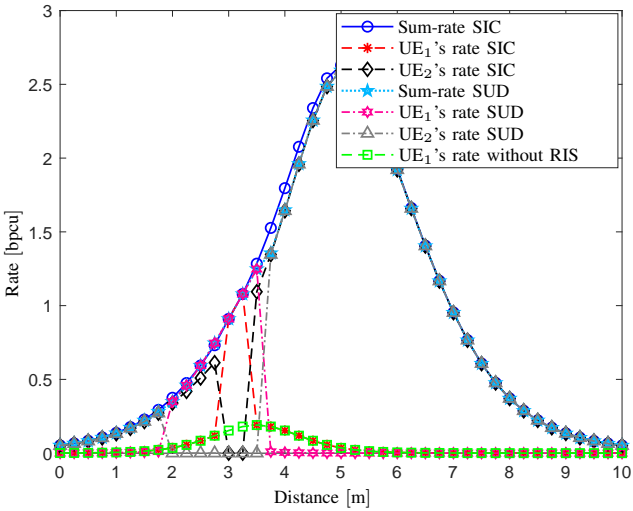}{
    \psfrag{SumRateSIC}{\hspace{-0.00cm}\scriptsize \text{Sum-rate SIC}}
    \psfrag{Rate1SIC}{\hspace{-0.00cm}\scriptsize \text{UE$_1$'s rate SIC}}
    \psfrag{Rate2SIC}{\hspace{-0.00cm}\scriptsize \text{UE$_2$'s rate SIC}}
    \psfrag{SumRateSUD}{\hspace{-0.00cm}\scriptsize \text{Sum-rate SUD}}
    \psfrag{Rate1SUD}{\hspace{-0.00cm}\scriptsize \text{UE$_1$'s rate SUD}}
    \psfrag{Rate2SUD}{\hspace{-0.00cm}\scriptsize \text{UE$_2$'s rate SUD}}
    \psfrag{Rate1WithoutRIS12345678910}{\hspace{-0.00cm}\scriptsize \text{UE$_1$'s rate without RIS}}
    \psfrag{Rate}{\hspace{-0.00cm}\scriptsize \text{Rate $[\text{bpcu}]$}}
    \psfrag{Distance}{\hspace{-0.35cm}\scriptsize \text{Distance $[\text{m}]$}}
    }

   \caption{User rates-- versus access point's $x$ position in room 1. UE$_1$ and UE$_2$ are fixed at $x=3.5\,m$ and $x=6\,m$ respectively.}
\label{Fig:Fig.10}
\end{figure}

Fig.~\ref{Fig:Fig.10} depicts sum-rate versus AP position in $x$ direction for a scenario wherein UE$_1$ and UE$_2$ are fixed at $x=3.5\,m$ and $x=6\,m$ respectively. As the AP moves toward RIS, sum-rate increases and at the location of RIS, maximum sum-rate is achieved. The reason is that as the AP reaches RIS, $d_{i,AP}$ in (\ref{Eq:Eq3}) decreases, therefore reflection and transmission channel coefficient, $H_i^r$ and $H_i^t$, increase and lead to improvement in sum-rate. For both SUD and SIC schemes, this trend is the same.
%%=============================================================%%
%===>Rate-AP Distance: UE1 & UE2 fixed<===%

%%=============================================================%%

\subsubsection{Mode-Switching}
\begin{figure}[ht]
    \centering
    \pstool[scale=0.55]{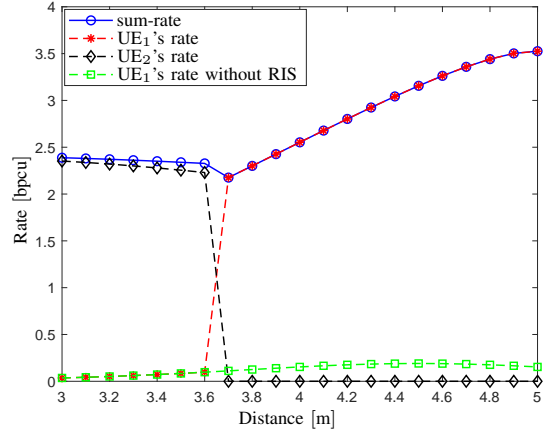}{
    \psfrag{SumRate}{\hspace{-0.00cm}\scriptsize \text{sum-rate}}
    \psfrag{Rate1}{\hspace{-0.00cm}\scriptsize \text{UE$_1$'s rate}}
    \psfrag{Rate2}{\hspace{-0.00cm}\scriptsize \text{UE$_2$'s rate}}
    \psfrag{Rate1Without RIS12345678910}{\hspace{-0.00cm}\scriptsize \text{UE$_1$'s rate without RIS}}
    \psfrag{Rate}{\hspace{-0.00cm}\scriptsize \text{Rate $[\text{bpcu}]$}}
    \psfrag{Distance}{\hspace{-0.35cm}\scriptsize \text{Distance $[\text{m}]$}}
    }
    \caption{User rates-- versus UE$_1$'s $x$ position in room 1 for mode-switching utilizing SIC at the AP. UE$_2$ is fixed at $x=6\,m$.}
    \label{Fig:Fig.11}
\end{figure}

Fig.~\ref{Fig:Fig.11} depicts user rates versus UE$_1$ distance for mode-switching and SIC case which is exactly the same as energy-splitting case shown in Fig.~\ref{Fig:Fig.6} (a). As mentioned before in Section \ref{Sec:Sec4}, since $\max f(\boldsymbol{\beta^r})=\max\big(f(0),f(1)\big)$  energy-splitting and mode-switching has the same performance in terms of the sum-rate.
%%=============================================================%%
%===>Mode Switching<===%
%%=============================================================%%
%%=====================Convergence Analysis====================%%
\subsubsection{Convergence Analysis of SPCA}
\begin{figure}[ht]
    \centering
    \pstool[scale=0.55]{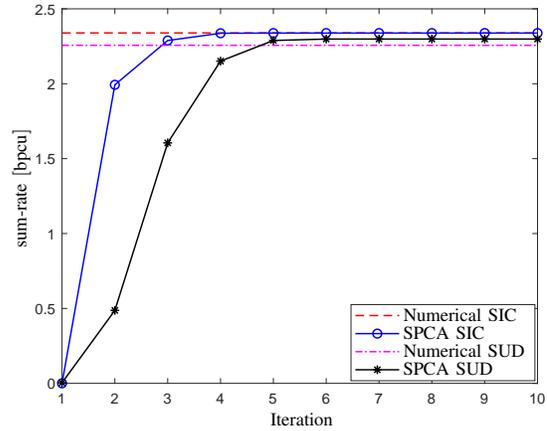}
    {
    \psfrag{Rate}{\hspace{-0.00cm}\scriptsize \text{sum-rate $[\text{bpcu}]$}}
    \psfrag{Iter}{\hspace{-0.25cm}\scriptsize \text{Iteration}}
    \psfrag{Numerical SUD12345}{\hspace{-0.00cm}\scriptsize \text{Numerical SUD}}
    \psfrag{SPCA SUD}{\hspace{-0.00cm}\scriptsize \text{SPCA SUD}}
    \psfrag{Numerical SIC}{\hspace{-0.00cm}\scriptsize \text{Numerical SIC}}
    \psfrag{SPCA SIC}{\hspace{-0.00cm}\scriptsize \text{SPCA SIC}}
    }
    \caption{Optimized sum-rate-- versus iteration of SPCA algorithm with initial $\theta^{(0)}=100$. UE$_1$ and UE$_2$ are fixed at $x=3.5\,m$ and $x=6\,m$ respectively.}
    \label{Fig:Fig.14}
\end{figure}

Fig. ~\ref{Fig:Fig.14} depicts  sum-rate versus number of iterations of SPCA algorithm with initial $\theta^{(0)}=100$. As it can be seen, the algorithm has converged to an optimum value in the forth and fifth iterations for SIC and SUD respectively. However, in the case of SUD, optimum value of SPCA is slightly higher than numerical result, because we have used a convex upper bound instead of (\ref{Eq:C1P5}). It is also worth noting that the computational complexity of SPCA algorithm is in polynomial time, since it has to solve several convex optimization problems using interior point methods \cite{ShenRateMaximization}.
%========>Convergence Analysis<========%
%%=============================================================%%
%%====> 6. Conclusion <===%%
\section{Conclusion} \label{Sec:Sec6}
In this work, a STAR-RIS based optical wireless uplink communications is introduced to investigate the effect of STAR-RIS on the performance of system’s sum-rate. Specifically, users rates have been derived, assuming that the AP uses SUD or SIC for detection. Moreover, two sum-rate optimization problem have been formulated for energy-splitting and mode-switching STAR-RIS, followed by an SPCA method for solving the problems.  In addition,  it has been deduce that the two operations of RIS have the same performance by showing that the maximum of objective function will always occur at $\boldsymbol{\beta^r},\boldsymbol{\beta^t}=0$ or $1$. At the end, the results have been verified using numerical simulation, considering SIC, SUD, time sharing and max-min fairness of users and analyzing convergence of SPCA algorithm.
\newpage
%%====> Appendixes <===%%
\appendices
%%====> References <===%%
\bibliographystyle{IEEEtran}
\bibliography{References.bib}

\end{document}